\newcounter{myctr}
\def\myitem{\refstepcounter{myctr}\bibfont\noindent\ifnum\themyctr>9\else\phantom{0}\fi\hangindent17pt\themyctr.\enskip}

\newcommand{\Tr}{\mbox{Tr}}
\def\beq{\begin{equation}}
\def\eeq{\end{equation}}
\def\bea{\begin{eqnarray}}
\def\eea{\end{eqnarray}}
\providecommand{\openone}{\leavevmode\hbox{\small1\kern-3.8pt\normalsize1}}

\documentclass{ws-ijqi}

\begin{document}

\markboth{Filippo Caruso, Vittorio Giovannetti} {A new approach to
characterize qubit channels}

%%%%%%%%%%%%%%%%%%%%% Publisher's Area please ignore %%%%%%%%%%%%%%
\catchline{}{}{}{}{}
%%%%%%%%%%%%%%%%%%%%%%%%%%%%%%%%%%%%%%%%%%%%%%%%%%%%%%%%%%%%%%%%%%%

\title{A NEW APPROACH TO CHARACTERIZE QUBIT CHANNELS}

\author{FILIPPO CARUSO}

\address{NEST CNR-INFM \& Scuola Normale Superiore, Piazza
dei Cavalieri 7 \\ Pisa, I-56126, Italy\\
filippo.caruso@sns.it}

\author{VITTORIO GIOVANNETTI}

\address{NEST CNR-INFM \& Scuola Normale Superiore, Piazza
dei Cavalieri 7 \\ Pisa, I-56126, Italy\\
vittorio.giovannetti@sns.it}

\maketitle

%\begin{history}
%\received{Day Month Year}
%\revised{Day Month Year}
%\accepted{Day Month Year}
%\comby{(xxxxxxxxxx)}
%\end{history}

\begin{abstract}
We analyze qubit channels by exploiting the possibility of
representing two-level quantum systems in terms of characteristic
functions.  To do so, we use functions of non-commuting variables
(Grassmann variables), defined in terms of generalized
displacement operators, following an approach which resemble the
one adopted for continuous--variable (Bosonic) systems. It allows
us to introduce the notion of qubit Gaussian channels and to show
that they share similar properties with the corresponding
continuous--variable counterpart. Some examples of qubit channels
are investigated using this  approach.
\end{abstract}


\section{Introduction}

In quantum information the transmission of messages through a
noisy environment is described by quantum channels. They are
mathematically represented by completely positive
trace--preserving maps acting on the set of the density operators
of the quantum system carrying
information\cite{nielsen,SHOR,KEYL,COVER}. For this reason the
characterization and classification of such maps have attracted a
lot of interest in recent years\cite{LIST}. The majority of the
results obtained so far relate to two specific classes of
channels, i.e. the qubit
channels\cite{RUSKAI,ruskai2,KINGUNITAL,DEVSHOR,KOLDAN,KINGDEP,QC,wolf}
and the Bosonic Gaussian
channels\cite{HOLEVOBOOK,HW,GL,SEW,CAVES,LOSS,NOSTRO,NOSTRO1,WOLF1,HOLEVONEW,WOLFLAST,REV1,GLAUBEROPT,WMILBURN}.

A quantum channel ${\cal N}$ can be
described\cite{HPPI,LINDBLAD,STINE} as a unitary interaction
between the information carrier system $S$ and an external
environment $E$, prepared in some fixed (generally) {\em mixed}
state $\rho_E$, i.e., ${\cal N}(\rho) = \mbox{Tr}_E[ U(\rho
\otimes \rho_E) U^\dag]$, where $\mbox{Tr}_E [ ... ]$ is the
partial trace over the environment and $U$ is a unitary operator
in the composite Hilbert space ${\cal H}_S \otimes {\cal H}_E$.
Following Refs. \refcite{NOSTRO,NOSTRO1} we generalize the  notion
of {\em complementary channel} ${\cal N}_{\text{com}}$ of $\cal N$
of Refs. \refcite{DEVSHOR,HOLEVOREP,KING}, by introducing the {\em
weakly complementary} channel ${\tilde{\cal N}}$ that maps the
input state $\rho$ of the carrier into the state of the
environment $E$ after the interaction with $S$, i.e., $\tilde{\cal
N}(\rho) = \mbox{Tr}_S[ U (\rho \otimes \rho_E) U^\dag]$. The
channel $\tilde{\cal N}$ describes, in some sense, the quantum
information lost in the environment due to the noise effects. The
quantum channel ${\cal N}$ is then called\cite{NOSTRO,NOSTRO1}
weakly degradable when, without having access to the input state
of the carrier $\rho$,  one can recover the information lost in
the environment during the interaction with $E$ only applying a
local physical transformation to the final system state after the
interaction. In the opposite case, the channel is called
anti-degradable, i.e. the output system state can be obtained
acting locally only on the environmental output state; in this
case the quantum capacity can be proved to be equal to
zero\cite{NOSTRO}.

In a previous paper\cite{NOSTRO2} we establish a parallelism among
the qubit channels and the Bosonic Gaussian channels, through a
phase-space representation of two-level quantum
states\cite{CAHILL}. Therefore, we introduce the set of qubit
Gaussian channels and we study the weak-degradability properties.
Interestingly enough Gaussian maps share similar properties in
both the qubit and Bosonic case. Here we briefly review these
results and we apply them to some examples of qubit channels.

\section{Characteristic and Green functions for qubits}\label{s:charafer}

The characteristic function description of  a qubit can be
introduced by properly adapting the formalism
of Ref. \refcite{CAHILL} that generalizes the Bosonic phase space
description to Fermionic systems.
As shown in Ref.~\refcite{NOSTRO2} this allows one to define
the characteristic function $\chi( \xi)$ of a density matrix $\rho$ as
$\chi( \xi )\equiv \Tr
[\rho D(\xi)]$ where $D(\xi)$ is
the qubit displacement operator $D (\xi) \equiv \exp \left( \sigma_+ \xi - \xi^*
\sigma_- \right)$ with $\sigma_+=(\sigma_-)^\dag = |1\rangle \langle 0|$ and
$\xi$ and $\xi^*$ being a couple of
conjugate Grassmann variables\cite{GRASS}.
For a  generic density operator of the form
$\rho \equiv
 \left(\begin{array}{cc}
 p  & \gamma \\
 \gamma^* & 1-p  \end{array} \right)$
this yields  $ \chi(\xi) = 1 + (2p-1) \frac{\xi
\xi^*}{2}  + \gamma \xi - \gamma^* \xi^*$.

Now let us consider the action of a qubit channel ${\cal N}$ on an
input state represented by the density operator $\rho$. Using the
above definitions one can write\cite{NOSTRO2} the characteristic
function $\chi^\prime (\xi )$ associated with the output state
${\cal N} (\rho)$  as
 \bea
 \chi^\prime (\xi ) &=& \int
d^2\zeta  \; \chi(\zeta) \; G(\zeta,\xi) \; ,
 \eea
where  \bea G(\zeta,\xi) &=& \Tr \Big[ {\cal N}
\Big( \sigma_3 D(-\zeta) \Big ) D(\xi)\Big] \; , \label{green}
 \eea
is the Green function of the map ${\cal N}$ (in this expression
$\sigma_3$ is the third Pauli matrix). The Green function
representation~(\ref{green}) gives a complete description of the
channel. In particular let us consider the canonical form of qubit
channels of Ref.~\refcite{RUSKAI}, i.e. ${\cal N}(\rho)={\cal
N}\left( \frac{\openone + {\vec{r}} {\mathbf \cdot
\vec{\sigma}}}{2} \right) = \frac{\openone +   ({\vec{t}} + T
{\vec{r}}) \cdot \vec{\sigma} }{2}$ with ${\vec{t}} = ({t}_1,
{t}_2, {t}_3)$ being a real vector,
$\vec{\sigma}=\{\sigma_1,\sigma_2,\sigma_3\}$ a vector containing
the Pauli matrices, ${\vec{r}}$ the Bloch vector describing the
input state, and $T=diag(\lambda_1, \ \lambda_2, \ \lambda_3)$,
with the real coefficients $\lambda_{1,2,3}$ and $t_{1,2,3}$
satisfying certain conditions\cite{RUSKAI,ruskai2}. The
corresponding Green function is
 \bea
&& G(\zeta,\xi) = \delta^{(2)}
 \left(\zeta-\frac{\lambda_2+\lambda_1}{2} \xi
 - \frac{\lambda_2-\lambda_1}{2} \xi^* \right) \exp\left[-\frac{t_3}{2} \xi^* \xi \right]
 \nonumber \\
&&\quad +(\lambda_3-\lambda_1 \lambda_2)\xi \xi^*+\frac{t_1-i
t_2}{2} \zeta \zeta^*\xi -\frac{t_1+i t_2}{2}\zeta \zeta^*\xi^*
\;,  \label{green.qubit}
 \eea
with $\delta^{(2)}(\zeta)$ being the Dirac delta function associated
with the Grassmann variable $\zeta$\cite{GRASS}.
In analogy with the Bosonic case, a qubit Gaussian channel is
defined by having the Green function of the form
 \bea
G(\zeta,\xi) = \delta^{(2)} ( \zeta - a \xi - b \xi^*) \; \exp[- c
\xi^* \xi] \label{guass.cond} \;,
 \eea
with $a$, $b$ complex and $c$ real (see Ref. \refcite{NOSTRO2} for
 details). According to Eq. (\ref{green.qubit}) the Gaussian
maps are obtained for $\lambda_3=\lambda_1 \lambda_2$,
$t_1=t_2=0$. With a proper parametrization, the Green function of
qubit Gaussian channels can thus be written as
 \bea
G(\zeta,\xi) = \delta^{(2)} \left(\zeta - \xi \cos \theta \cos \phi +\xi^*
\sin \theta \sin \phi \right)\exp \left[ (2q-1) \;\frac{\cos (2
\theta) - \cos (2 \phi)}{4} \xi \xi^*
   \right]  \nonumber
 \eea
with $\theta$, $\phi$ in $[0,2\pi[$ and $q\in [0,1]$. It is worth pointing
out  that, analogously to what happens in the Bosonic case\cite{NOSTRO,NOSTRO1},
the qubit Gaussian channels defined here can always be
described as ``qubit-qubit'' channels, i.e. in terms of a unitary
interaction between a qubit system and a {\em single} (not
necessarily pure) qubit environment\cite{NOSTRO2}.
Pure environment is obtained when
$q=0$ or $q=1$, $\theta$ and $\phi$ generic. These channels have
been proved\cite{wolf,NOSTRO2} to be weakly degradable for
$\cos(2\theta)/\cos(2\phi) \geqslant 0$, and anti-degradable
otherwise. If $0<q<1$, the single qubit environment $E$ is
initially prepared in the mixed state $\rho_E \equiv q | 0
\rangle_E \langle 0 | + (1-q) | 1 \rangle_E \langle 1 |$, and the
channel is weakly degradable\cite{NOSTRO2} for
$\cos(2\theta)/\cos(2\phi) \geqslant 0$ and with null quantum
capacity otherwise.

\section{Some qubit channels}
\label{examples}

In the following we will show the Green function of some examples
of qubit quantum channels\cite{nielsen} and we will analyze their
weak-degradability properties.

\subsection{Bit flip or dephasing channel}

The bit flip (or dephasing channel) flips the state $|0\rangle$ to
$|1\rangle$ (and viceversa) with probability $1-s$. This map can
be obtained from the canonical form above by $t_1=t_2=t_3=0$,
$\lambda_1=1$, and $\lambda_2=\lambda_3=2 s - 1$; it is a
qubit-qubit map with pure environment (q=1). The relative Gaussian
Green function is:
 \bea
G(\zeta,\xi) &=& \delta^{(2)}
 \left(\zeta-s \xi
 - (s-1) \xi^* \right)\; .
 \eea
Observing that $\cos(2 \theta)=\frac{\lambda_3+t_3}{2}=s-1/2$,
$\cos(2 \phi)=\frac{\lambda_3-t_3}{2}=s-1/2$, and $\cos(2
\theta)/\cos(2 \phi)=1>0$, the bit flip channel is always weakly
degradable for any value of s.

\subsection{Phase flip channel}

The phase flip channel changes the phase of the state $|1\rangle$
with probability $1-s$; for instance,
$\frac{1}{2}(|0\rangle+|1\rangle)$ is mapped to
$\frac{1}{2}(|0\rangle-|1\rangle)$ with probability $1-s$ and with
probability $s$ it remains unchanged. This channel has the
following (canonical) parameters $t_1=t_2=t_3=0$, $\lambda_3=1$,
and $\lambda_1=\lambda_2=2 s - 1$ and the relative Green function
is not Gaussian, i.e.
 \bea
G(\zeta,\xi) &=& \delta^{(2)}
 \left(\zeta-(2 s -1) \xi \right)
 +4s(1-s)\xi \xi^* \;.
 \eea
Since the canonical form\cite{ruskai2} is uniquely determined only
up to unitary transformations, one can permute the $\lambda$s and
so the phase flip channel is unitarily equivalent to a bit-flip
channel. Therefore, the phase flip channel is not a Gaussian
channel but it is unitarily equivalent to a (weakly degradable)
Gaussian map.

\subsection{Bit-phase flip channel}

The bit-phase flip channel is a combination of a bit flip and a
phase flip channels\cite{nielsen}. Its Kraus operators
are:
 \bea
 A_0 &=& \sqrt{s} \ I = \sqrt{s} \ \left(\begin{array}{cc}
1 & 0 \\
0 & 1
\end{array}\right) \ , \ \ \ A_1 = \sqrt{1-s} \ \sigma_y = \sqrt{1-s}
\ \left(\begin{array}{cc}
0 & -i \\
i & 0
\end{array}\right).
 \eea
The phase-flip channel is obtained with the following parameters
$t_1=t_2=t_3=0$, $\lambda_2=1$, $\lambda_1=\lambda_3=2 s - 1$, and
so it is a qubit-qubit map with pure environment (q=1). Indeed,
$|t_3| = \sqrt{(1-\lambda_1^2) (1-\lambda_2^2)}=0$. The Green
function is so Gaussian, i.e.
 \bea
G(\zeta,\xi) &=& \delta^{(2)}
 \left(\zeta-s \xi
 - (1-s) \xi^* \right) \;.
 \eea
Observing that $\cos(2 \theta)/\cos(2 \phi)=1>0$, the bit-phase
flip channel is always weakly degradable for any value of s.

\subsection{Depolarizing channel}

The depolarizing channel represents an important kind of noise
evolution, in which the qubit is \textit{depolarized} (i.e.
replaced by the completely mixed state, $I/2$) with probability
$s$ and it is left untouched with probability $1-p$. Therefore,
the output state, i.e. $ {\cal N}(\rho)=\frac{s}{2} I + (1-s)
\rho$. In the canonical representation it is characterized by the
parameters $t_1=t_2=t_3=0$, $\lambda_1=\lambda_2=\lambda_3=1-s$,
and, since the Green function is not Gaussian, i.e.
 \bea
G(\zeta,\xi) = \delta^{(2)}
 \left(\zeta-(1-s) \xi \right) + s(1-s)\xi \xi^* \; ,
 \eea
we are not able to discuss its degradability properties in our
formalism.

\subsection{Amplitude damping channel}

Let us consider now a typical process of noise evolution in which
a quantum system losses its energy, well implemented by the
\emph{amplitude damping} channel. In the canonical form the
amplitude damping channel is given by $t_1=t_2=0$, $t_3=1-n$,
$\lambda_1=\lambda_2=\sqrt{n}$, and $\lambda_3=n$, where $1-n$ can
be thought of as the probability of losing the system energy.
Since $|t_3| = \sqrt{(1-\lambda_1^2) (1-\lambda_2^2)}=1-n$, it is
a qubit-qubit map with pure environment (q=1). The Gaussian Green
function has the form
 \bea
G(\zeta,\xi) &=& \delta^{(2)}
 \left(\zeta-\sqrt{n} \xi\right) \exp
 \left[-\frac{1-n}{2} \xi^* \xi \right] \;.
 \eea
Since $\cos(2 \theta)/\cos(2 \phi)=\frac{1}{2n-1}$, the amplitude
damping channel is weakly degradable for $n \geq 1/2$ and
anti-degradable for $n \leq 1/2$.

\subsection{Generalized amplitude damping channel}

Here we describe the effect of dissipation due to the presence of
an external environment at finite temperature. This quantum
operation, called generalized amplitude damping channel, can be
described by the following Kraus operators ($s \neq
1$)\cite{nielsen}:
 \bea
 A_0 &=& \sqrt{s} \ \left(\begin{array}{cc}
1 & 0 \\
0 & \sqrt{n}
\end{array}\right)\ , \ \ \  A_1 = \sqrt{s} \ \left(\begin{array}{cc}
0 & \sqrt{1-n} \\
0 & 0
\end{array}\right) \ ,\\
 A_2 &=& \sqrt{1-s} \ \left(\begin{array}{cc}
\sqrt{n} & 0 \\
0 & 1
\end{array}\right) \ , \ \ \ A_3 = \sqrt{1-s} \ \left(\begin{array}{cc}
0 & 0 \\
\sqrt{1-n} & 0
\end{array}\right) \nonumber \; .
 \eea
The generalized amplitude damping channel corresponds to
$t_1=t_2=0$, $t_3=(1-n) (2s-1)$, $\lambda_1=\lambda_2=\sqrt{n}$
and $\lambda_3=n$. The Gaussian Green function is
 \bea
G(\zeta,\xi) &=& \delta^{(2)}
 \left(\zeta-\sqrt{n} \xi \right) \exp
 \left[- (2 s-1)\frac{(1-n)}{2} \xi^* \xi \right] \;.
 \eea
It is a qubit-qubit map with mixed environment ($q \equiv s \neq
1$) and is weakly degradable for $n \geq 1/2$ and with null
quantum capacity for $n \leq 1/2$.

\section{Conclusions}\label{s:conclusion}

In this work we briefly review the interesting
parallelism\cite{NOSTRO2} among the qubit channels and the Bosonic
Gaussian channels, through a phase-space
representation\cite{CAHILL} in terms of generalized characteristic
functions for qubits. It allows one to define the qubit Gaussian
channels and to study their weak-degradability properties,
showing in a elegant way a strong analogy to what is obtained for
Bosonic Gaussian channels. Therefore, we use these results to
describe some examples of qubit channels, which play an important
role in quantum information science and its applications. We find
that not all qubit channels are Gaussian but only those
describable trough a noisy interaction between one qubit (for the
system) and one qubit (for the environment).

\section*{Acknowledgments}
This work was supported in part by the Centro di Ricerca Ennio De
Giorgi of the Scuola Normale Superiore of Pisa.

\end{document}